\documentclass[twocolumn,aps,,prl,reprint]{revtex4-2}

\usepackage[utf8]{inputenc}
\usepackage{epsf,graphicx}
\usepackage{natbib}
\usepackage{multirow}
\usepackage{amsmath,gensymb,amssymb}
\usepackage{siunitx}
\usepackage[bottom]{footmisc}
\usepackage{lipsum}
\usepackage{dcolumn}
\usepackage{xcolor}
\usepackage{upgreek}
\usepackage{float,ulem}

\usepackage{bm}
\usepackage{hyperref}
\hypersetup{
colorlinks=true, 
citecolor=magenta,
breaklinks=true, 
urlcolor= blue, 
linkcolor= blue, 
bookmarksopen=true, 
}
\usepackage{natbib}
\usepackage{color}
\usepackage[type={CC},modifier={by-nc-sa},version={4.0},]{doclicense}
\definecolor{linkcolor}{rgb}{0,0,0.6}

\begin{document}

\title{Soliton Dynamics over a Disordered Topography}	
	

\author{Guillaume Ricard}
\altaffiliation[Present address: ]{Faculty of Mechanical Engineering, Delft University of Technology, The Netherlands}
\affiliation{Universit\'e Paris Cité, CNRS, MSC, UMR 7057, F-75013 Paris, France}

\author{Eric Falcon}
\email{eric.falcon@u-paris.fr}
\affiliation{Universit\'e Paris Cité, CNRS, MSC, UMR 7057, F-75013 Paris, France}

\begin{abstract}
We report on the dynamics of a soliton propagating on the surface of a fluid in a 4-m-long canal with a random or periodic bottom topography. Using a full space-and-time resolved wavefield measurement, we evidence, for the first time experimentally, how the soliton is affected by the disorder, in the context of Anderson localization, and how localization depends on nonlinearity. For weak soliton amplitudes, the localization length is found in quantitative agreement with a linear shallow-water theory. For higher amplitudes, this spatial attenuation of the soliton amplitude is found to be enhanced. Behind the leading soliton slowed down by the topography, different experimentally unreported dynamics occur: Fission into backward and forward nondispersive pulses for the periodic case, and scattering into dispersive waves for the random case. Our findings open doors to potential applications regarding ocean coastal protection against large-amplitude waves.
\end{abstract}

\maketitle


\textit{Introduction.---}
Wave propagation in nonhomogeneous media leads to astonishing phenomena such as Bragg reflection~\cite{Bragg1913} and Anderson localization~\cite{Anderson1958}, first discovered in solid-state physics. These linear phenomena, based on interferences between multiply-scattered waves by the changes in the spatial property of the medium, lead to a significant attenuation of the incident waves. They occur in almost all domains involving a linear wave field~\cite{Lagendijk2009} from cold atoms~\cite{Roati2008} to classical waves (such as acoustics~\cite{Hodges1982}, optics~\cite{StorzerPRL06,PertschPRL04,Lahini2008}, electrical waves~\cite{Akkermans1984}), as well as for gravitational waves~\cite{Rothstein2013} or random time-varying media~\cite{Sharabi2021,Apffel2022}. For gravity waves on the surface of a fluid, Anderson localization has been observed using a random bathymetry~\cite{Belzons1987,Ricard2024} and Bragg reflection using a periodic one~\cite{Davies1984,Guazzelli1992}.

Such phenomena occur mainly for linear waves, and their persistence for nonlinear waves is a long-standing debate. For initially monochromatic waves, it has been experimentally shown that nonlinearities arising from the increase of their amplitude enhance Anderson localization~\cite{Mckenna1992,PertschPRL04,Lahini2008,Ricard2024} as suggested theoretically~\cite{Frohlich1986}. For nonlinear pulses and solitons, Anderson localization is theoretically predicted to be destroyed for the nonlinear Schr{\"o}dinger equation~\cite{Kivshar1990,PikovskyPRL08,Ivanchenko2011} and disordered anharmonic chains~\cite{Li1988,Bourbonnais1990}, but to persist for the Korteweg-deVries (KdV) equation~\cite{Nakoulima05}. However, to our knowledge, the effects of pulse nonlinearity on Anderson localization remain experimentally elusive. Previous experiments performed with a superfluid helium film failed to achieve sufficient amplitude of the nonlinear pulse for its width to be smaller than the localization length~\cite{Hopkins1996}. 

Here, we experimentally report how a KdV soliton propagating on the surface of a fluid in a canal is affected by a random or periodic bottom topography. The spatiotemporal dynamics of such elevation solitons are measured by five cameras regularly spaced laterally along a 4-m-long canal. In particular, we show experimentally that the pulse localization is enhanced by nonlinearity as in the case of large-amplitude sinusoidal waves~\cite{Ricard2024}, but to a lesser extent. For weak soliton amplitudes, the localization length is found to be in quantitative agreement with a linear shallow-water theory~\cite{Nakoulima05}. 
Different dynamics are also highlighted behind the leading attenuated soliton:  Fission into slower, nondispersive pulses for the periodic lattice, and scattering into dispersive waves for the random case. Soliton dynamics over a random topography in shallow water has been predicted theoretically for a disorder scale, $L$, smaller than the soliton width $l$~\cite{Rosales83,Nachbin2003,Mei2004,Garnier2007}. Here, we experimentally focus on the opposite case ($L\gtrsim l$) and compare it to the corresponding predictions in the context of Anderson localization~\cite{Nakoulima05}. Note that a few experimental and numerical studies have been performed for a soliton propagating over an isolated immersed obstacle~\cite{Santos1985,Losada1997,Wu2012,Chao2014,Fu2024} or a continuous step~\cite{Pelinovsky2010,Losada1989}, but not over a lattice of immersed obstacles, despite potential applications regarding the ocean coastal protection against large-amplitude waves such as tsunamis~\cite{Pelinovsky1996,Zhang2012}. 

\begin{figure}[t!]
    \centering
    \hspace{-1cm}
    \includegraphics[width=1.1\linewidth]{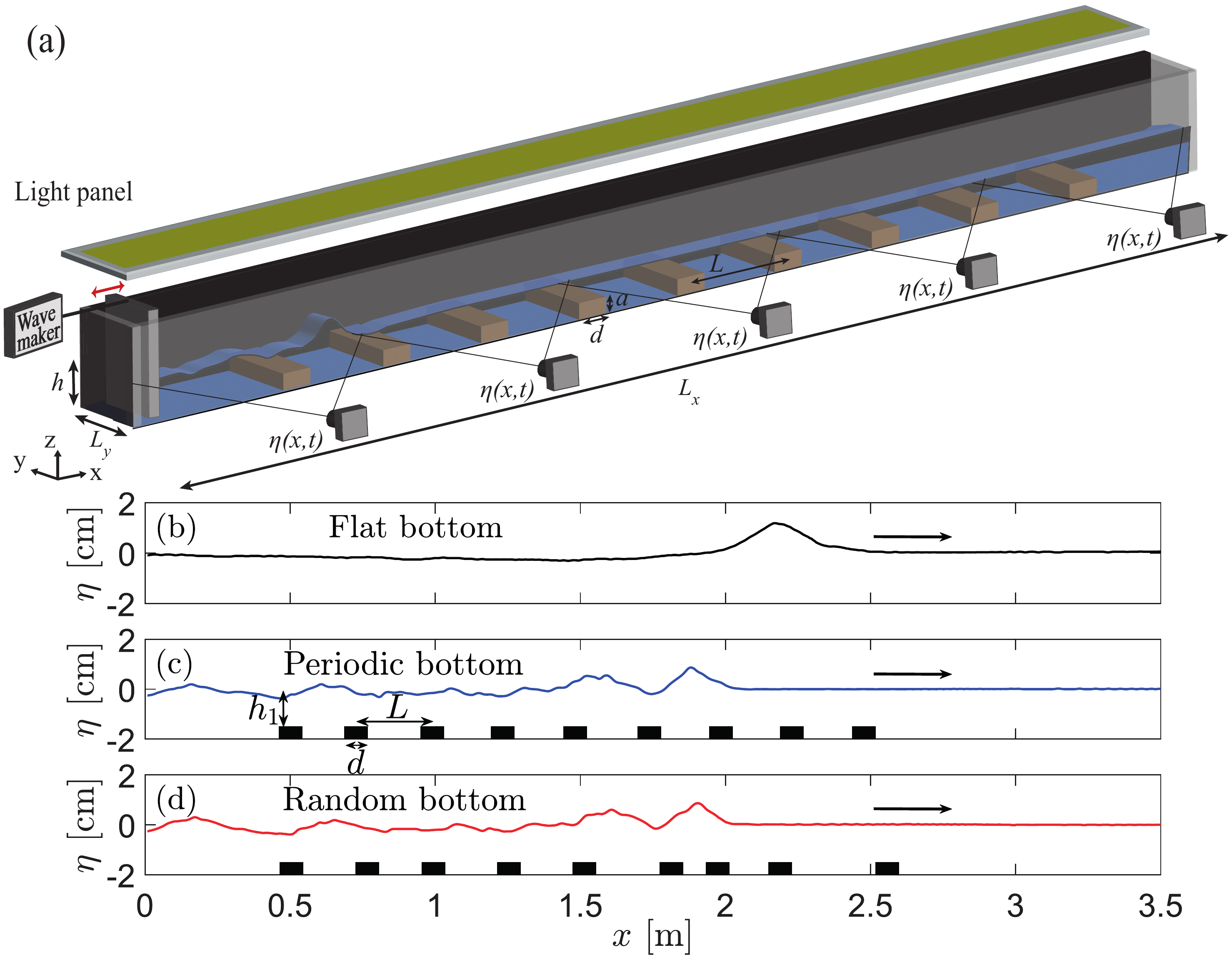} 
    \caption{(a) Experimental setup to study soliton propagation in a 4-m canal over a random or periodic bathymetry. $a=4$ cm, $d=8$ cm, $L=25$ cm, and $h=5.5$ cm. Typical wave amplitude $\eta(x)$ for a (b) flat, (c) periodic, and (d) random bottom ($L=25+\delta$~cm) at $t=2.5$ s after the pulse generation ($A_0=1.5$ cm). Black rectangles indicate the bar locations.}
    \label{setup}
\end{figure}

\textit{Experimental setup and soliton generation.---}
The experimental setup is represented in Fig.~\ref{setup}. It consists of a transparent canal of length $L_x=4$~m and of width $L_y=18$~cm. The canal is filled with water to a depth $h=5.5$~cm. At one end, a linear motor drives impulsionally a paddle to generate pulse-like surface waves. To do so, we prescribe a horizontal displacement of the wave maker, $\alpha \tanh(t/\tau^*)$, with $\alpha$ its amplitude and $\tau^*$ its characteristic time. The parameters $\alpha$ and $\tau^*$ are chosen following the method described in~\cite{Guizien2002} to generate a soliton with a chosen amplitude $A_0$. The generated soliton is expected to be a solution of the KdV equation~\cite{KdV1895}. Its amplitude $A_0$ (before the lattice), its theoretical width $l=\sqrt{4h^3/(3A_0)}$ and supersonic velocity $c=c_0[1+A_0/(2h)]$ (with $c_0=\sqrt{gh}$ the phase speed of linear waves) are hence directly linked with each other (see~\cite{Remoissenet1994,FalconPRL2002} and Supplemental Material~\cite{SuppMat}). We will use different experimental amplitudes $A_0\in[0.25, 2.7]$~cm leading to widths $l\in[9,30]$~cm, nonlinear parameters $\epsilon=A_0/h\in[0.05, 0.5]$, and dispersive parameters $\mu=(h/l)^2\in[0.03, 0.37]$, both smaller than 1 and of the same order of magnitude to balance weak nonlinear and dispersive effects within an arbitrary depth fluid~\cite{Remoissenet1994,FalconPRL2002}. In the flat bottom case, the solitonic shape is first verified experimentally by fitting the experimental pulse with the KdV solution with no adjustable parameter once $A_0$ is fixed (see Supplemental Material~\cite{SuppMat}). We find that the taller the soliton, the better the agreement as a cavity is generated behind the leading pulse due to mass conservation. We will use the term soliton afterward, remembering that it is not fully true for the smallest amplitudes. The soliton propagates along the $x$ axis and is reflected at the end wall of the canal. 

The spatiotemporal measurement of the surface elevation $\eta(x,t)$ is made by the use of five cameras (Basler, 20 fps, 6 Mpixel) regularly spaced laterally along the canal [see Fig.~\ref{setup}], each filming an 86~cm side view with an angle of 4$^\circ$ to the horizontal, following the method described in~\cite{Redor2020}. The surface is illuminated from the top to generate a strong contrast between the water surface and a black lateral background. The horizontal and vertical resolutions are 0.28~mm/pixel. The five cameras are time synchronized to film simultaneously the water surface along the canal to rebuild the whole wavefield. 
To create a spatially-dependent bathymetry, $N=9$ rectangular aluminum bars of width $d=8$~cm and height $a=4$~cm are placed along the bottom of the canal. Hence, the water depth changes along the canal accordingly: $h=5.5$~cm between the bars and $h_1=h-a=1.5$~cm over each bar. For periodic bathymetry, the bars are located regularly every $L=25$~cm. The random bathymetry is created from the periodic lattice by moving each bar by a length $\delta$ chosen randomly and uniformly in the range $\delta \in [-\kappa L/2, \kappa L/2]$ for each bar. $\kappa$ thus quantifies the level of disorder from 0 (periodic case) to 1 (fully random). The values of $L$, $N$ and $\kappa=0$ or 1 are fixed. The spatial period of the lattice $L$ is larger than or of the order of the typical soliton width $l$ as $l/L\in[0.36,1.2]$, and we restrict our study to the short-obstacle case discussed theoretically~\cite{Nakoulima05}. The value of the obstacle height $a$ has been chosen to respect the short-obstacle approximation and to make experimental measurements possible within our 4-m-long canal. 

\textit{Space-time evolution and wave spectrum.---}
Examples of solitons traveling along the canal are shown in Fig.~\ref{setup} for a (b) flat, (c) periodic, and (d) random bottom (see also movies {\it soliton.mp4} in Supplemental Material~\cite{SuppMat}). We first observe that the bathymetry implies fissions of the main soliton generating many waves behind it. This dispersion involves a significant slowdown of the soliton compared to the flat case as its amplitude decreases. However, no difference emerges, at first sight, between the periodic and the random lattices.

\begin{figure*}[t!]
    \centering
    \includegraphics[width=\linewidth]{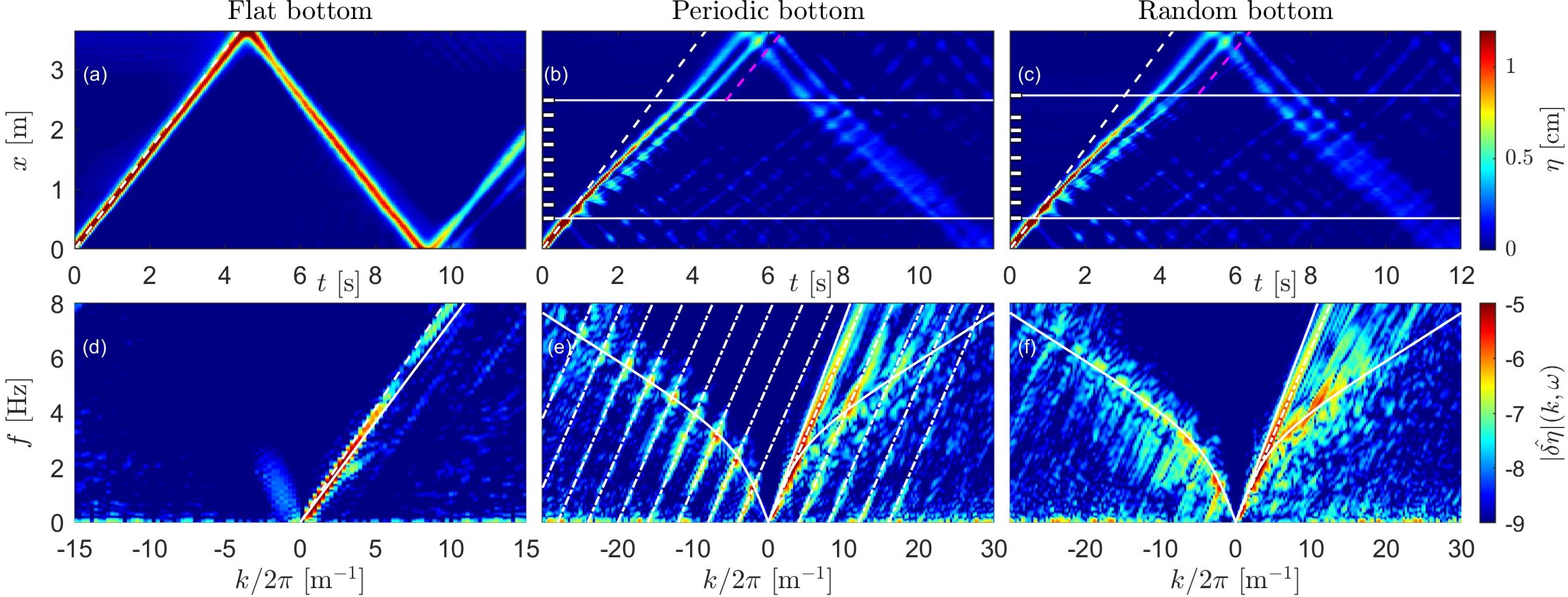} 
    \caption{Space-time evolution $\eta(x,t)$ of a soliton over a (a) flat, (b) periodic, and (c) random bottom ($A_0=1.5$ cm, $\epsilon= 0.3$). Dashed-white line marks $x=ct$ for the flat bottom. Magenta dashed line marks $x=c_ft$ with $c_f$ the velocity based on the amplitude after the bathymetry. White rectangles indicate the bar locations of the lattice delimited by the horizontal solid lines. Linear colorbar. Corresponding spatiotemporal spectrum $|\hat{\delta\eta}(k,\omega)|$ for a (d) flat, (e) periodic, and (f) random bottom (first trip only). Dashed line: $\omega=ck$ for the flat bottom. Solid-straight line: $\omega=c_0k$ with $c_0=\sqrt{gh}$. Dash-dotted lines $\omega=c^*k+k_n$ with $k_n=(2\pi n/L)$. Solid-curved lines: $\omega^2=gk\tanh(kh)$ the linear wave dispersion relation. Log colorbar.  
    }
    \label{spatio_temp}
\end{figure*}

We plot in Fig.~\ref{spatio_temp}(a), (b), and (c) the space-time evolution $\eta(x,t)$ of the soliton for the flat, periodic, and random case, respectively. We also plot in Fig.~\ref{spatio_temp}(d), (e), and (f) the corresponding spatiotemporal spectra $|\hat{\delta\eta}(k,\omega)|$ of the surface wavefield difference $\delta\eta(x,t)=\eta(x+dx)-\eta(x)$, with $dx$ the spatial resolution, taking only the first trip into account (i.e., before end reflection). Note that computing the spectrum of the surface elevation $|\hat{\eta}(k,\omega)|$ provides similar but noisier results.

For the flat bottom [Fig.~\ref{spatio_temp}(a) and (d)], a KdV soliton is observed, as expected. It travels along the canal with an almost constant velocity, predicted by $c=c_0[1+A_0/(2h)]$ (dashed lines), and is reflected at the end wall (at $t\approx4.5$~s), before going backward [see Fig.~\ref{spatio_temp}(a)]. The spatiotemporal spectrum of its first trip in Fig.~\ref{spatio_temp}(d) shows that the wave energy is spread along a straight line also characteristic of the constant velocity $c$ (see dashed lines) that is larger than the linear wave speed $c_0=\sqrt{gh}$ (solid line), meaning that this elevation soliton is indeed supersonic. A slight decrease of the soliton velocity occurs at the canal end because of a slight decrease in its amplitude due to viscous effects (see below).

For the periodic and random cases, the spatiotemporal evolutions of the soliton shown in Fig.~\ref{spatio_temp}(b) and (c) appear similar. The bathymetry strongly slows down the incident soliton, generating many waves behind it (see patterns between the two horizontal white lines delimiting the lattice). Reflections of the incident soliton at each step (see spots along a second straight line) lead to backward reflected waves. After the lattice, two main solitons, of reduced amplitude $A_f\approx0.4$~cm and velocity $c_f=c_0[1+A_f/(2h)]$, persist (see magenta dashed lines). However, despite strong similarities between the space-time evolutions, the periodic and random cases lead to very different spatiotemporal spectra [see Fig.~\ref{spatio_temp}(e) and (f)]. Indeed, for the periodic bottom, wave energy is spread along several straight lines in the ($k$, $\omega$) space of Fig.~\ref{spatio_temp}(e) (dash-dotted lines) as a signature of solitary waves. These lines have all the same slope and occur at each spatial period of the periodic lattice (i.e., $k_n=2\pi n/L$ with $n\in \mathbb{Z}$) showing that the incident soliton undergoes fission at each step. This periodic pattern is remarkable as it survives for nonlinear pulses compared to the spectrum periodicity of linear sine waves over a bathymetry~\cite{Ricard2024}. These solitonic structures are mostly present for $k<0$ corresponding to backward waves induced by reflections at each step. Additionally, a secondary forward solitary wave is observed for $k>0$ (red spot) corresponding to a reflected wave that has been reflected again in the forward direction. Such forward and backward waves have the same velocity, independent of $k$, and are well described by $\omega=c^*k+k_n$ (see dash-dotted lines) with $c^*=\sqrt{gh^*}$ an effective velocity based on the mean depth $h^*=0.65h+0.35h_1=4.1$~cm as $65\%$ of the bathymetry has a depth of $h$ and $35\%$ a depth of $h_1$. This means that the incident soliton breaks into nondispersive pulses with a mean-depth-based effective velocity smaller than $c_0=\sqrt{gh}$ the linear wave speed in the flat case (solid straight line). Thus, the topography implies a loss of the usual solitonic velocity $c=c_0[1+A_0/(2h)]$ observed only before and after the lattice. 

The spectrum for the random case in Fig.~\ref{spatio_temp}(f) is different. No solitonic wave emerges at each reflection of the lattice step. Instead, most of the energy (excluding the incident soliton one) is present along the classical linear dispersion relation $\omega^2=gk\tanh(kh)$, meaning that mostly dispersive waves are generated over the bathymetry. Despite similar spatiotemporal wave evolutions, a strong difference in physics is hence evidenced between the periodic and random cases.

\begin{figure}[t!]
    \centering
    \includegraphics[width=1\linewidth]{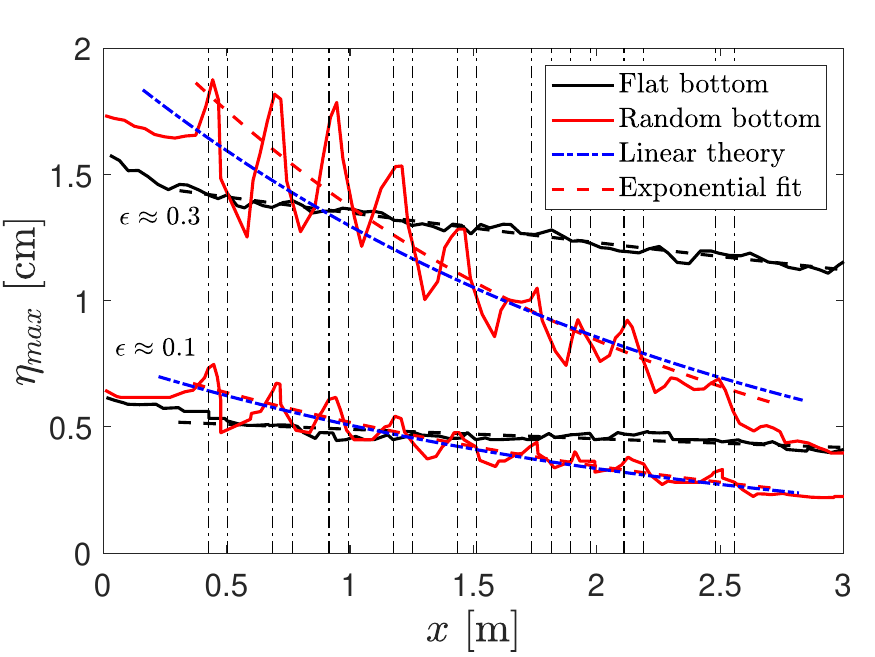}
    \caption{Evolution of the soliton maximum amplitude along the canal for (black) flat and (red) random bottoms, and two different initial amplitudes ($\epsilon \approx 0.1$ and $0.3$). Vertical dash-dotted lines represent the bar locations of the random case. Dashed lines: best exponential fits $\eta_{max}=\eta_0\exp{(-x/\xi)}$. Blue dash-dotted lines: theoretical evolutions of $\eta_{max}$ from Eq.~\eqref{loca_th_eq} and Eq.~\eqref{xi_th_eq}.}
    \label{etamax}
\end{figure}

\textit{Anderson localization and solitons.---}
As dissipation has the same signature as Anderson localization of multiple-scattered waves by disorder [i.e., exponential spatial decay of the amplitude, see Eq.~\eqref{loca_th_eq}], we first check that dissipation is negligible. To quantify it, we plot in Fig.~\ref{etamax} (black solid lines) the soliton maximum amplitude $\eta_{max}(x)$ along the flat bottom canal for two different initial amplitudes $A_0$ (i.e., $\epsilon \approx 0.1$ and $0.3$). $\eta_{max}(x)$ only slightly decreases and is well fitted by an exponential spatial decay $\eta_{max}(x)=\eta_0\exp{(-x/l_d)}$ (black dashed lines) with $l_d\approx 12\pm3$~m. This dissipative length $l_d$ is independent of $\epsilon$ and much longer than the canal length $L_x$, indicating that viscous effects are almost negligible.

\begin{figure}[t!]
    \centering
    \includegraphics[width=1\linewidth, height=6.5cm]{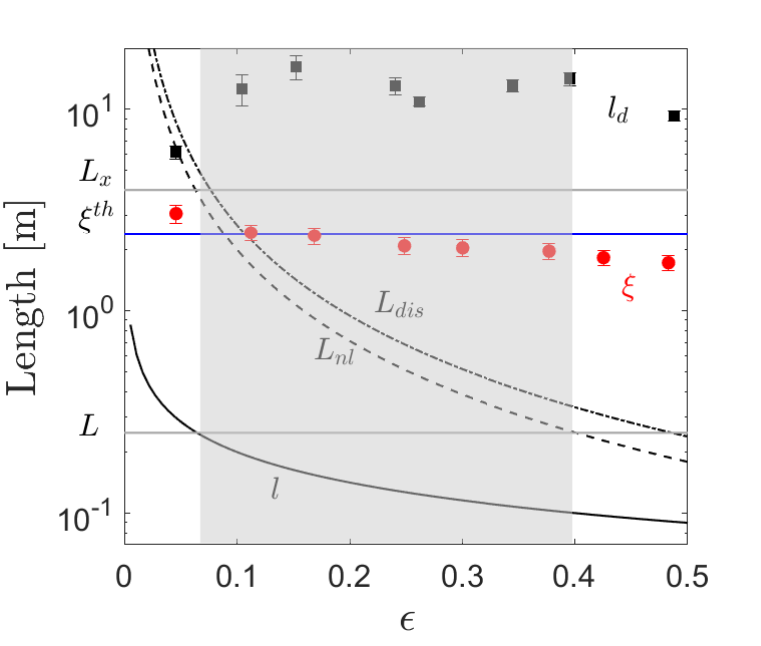}
    \caption{Localization lengths versus nonlinearity $\epsilon$: experiments (red bullets), linear theory $\xi^{th}$ (blue line) of Eq.~\eqref{xi_th_eq} valid within the gray area. Random bottom. Solid black line: theoretical soliton width $l$. Dashed black line: theoretical nonlinear length $L_{nl}=l/\epsilon$. Dash-dotted black line: dispersion length $L_{dis}=l/\mu=4L_{nl}/3$. Experimental dissipative length $l_d$ (black squares). Horizontal lines (from top to bottom): canal length $L_x$, theoretical localization length $\xi^{th}$, disorder scale $L$.  Errorbars are from the exponential fit errors (Fig.~\ref{etamax}).}
    \label{xiep}
\end{figure}

For the random bottom, the soliton amplitude $\eta_{max}(x)$ is much more altered by disorder as shown in Fig.~\ref{etamax} (red solid lines). Fluctuations of $\eta_{max}(x)$ are also observed at each random step location as a consequence of the generation of scattered waves. Despite the presence of such fluctuations, $\eta_{max}(x)$ can be fitted with an exponential spatial decay as, 
\begin{equation}
    \eta(x)=\eta_0\exp{(-x/\xi)}.
    \label{loca_th_eq}
\end{equation}
where $\xi$ is the Anderson localization length~\cite{Anderson1958}. The corresponding values of $\xi$ are then reported in Fig.~\ref{xiep} for different initial soliton amplitudes $A_0$, i.e., different nonlinearities~$\epsilon$. The localization length is found to decrease with the soliton amplitude. 
This means that Anderson localization is enhanced by nonlinearity as for the initially monochromatic wave case~\cite{Ricard2024}. The experimental values of $\xi$ are now compared with their theoretical values $\xi^{th}$ from a linear shallow-water theory of a soliton propagating over a bathymetry as~\cite{Nakoulima05}
\begin{equation}
    \xi^{th}=L/\ln{\left[\left(1+\sqrt{h/h_1}\right)^2/\left( 4\sqrt{h/h_1}\right)\right]}
    \label{xi_th_eq}
\end{equation}
that depends on the bathymetry features ($L$ and $h/h_1$) leading to $\xi^{th}=2.4$~m for our experimental parameters. The theoretical exponential attenuation of $\eta_{max}(x)$ from Eq.~\eqref{loca_th_eq} using $\xi^{th}$ from Eq.~\eqref{xi_th_eq} are then plotted in Fig.~\ref{etamax} in dash-dotted lines and well described the experiments for weak enough forcing. In Fig.~\ref{xiep}, this theoretical linear model $\xi^{th}$ from Eq.~\eqref{xi_th_eq} agrees with experiments for relative weak nonlinearities $\epsilon \in[0.1, 0.2]$. For stronger nonlinearity, a departure is observed as other phenomena such as vortices or capillarity are probably significant and are not captured by this linear model. The latter is derived by assuming relatively short obstacles, i.e., $l\lesssim L\ll L_{nl}\mathrm{,}\ L_{dis}$ with $L_{nl}\sim l/\epsilon=lh/A_0$ the nonlinear length and $L_{dis}\sim l/\mu =l^3/h^2$ the dispersive length~\cite{Nakoulima05}. These conditions are valid experimentally as shown in the gray area of Fig.~\ref{xiep} except for the smallest pulse $\epsilon \sim 0.05$ as the condition $l < L$ is not respected and its solitonic shape is not reached (see Supplemental Material~\cite{SuppMat}), explaining thus the departure with the theory. Similarly, the condition $L\ll L_{nl}\mathrm{,}\ L_{dis}$ is no longer valid for the two tallest pulses (see Fig.~\ref{xiep}). To sum up, one has the relevant length scales as $l\lesssim L\ll L_{nl}, L_{dis}\lesssim \xi < L_x \ll l_d$, as expected for solitons to be localized, with $l$ the soliton width, $L=0.25$ m the disorder scale, $\xi$ the localization length, $L_x=4$ m the system size, and $l_d\sim 12$ m the dissipative length. These conditions are well valid experimentally for most probed soliton amplitudes (see Fig.~\ref{xiep}).

\textit{Conclusion.---}
We have reported on the experimental dynamics of KdV elevation solitons propagating on a fluid surface over a flat, periodic, or random bathymetry in a canal. Using a spatiotemporal measurement, we have evidenced, for the first time experimentally, how a soliton is impacted by disorder in the context of Anderson localization. For weak soliton amplitudes, the localization length is found in agreement with a shallow-water theory~\cite{Nakoulima05}. We also found that localization is experimentally enhanced by nonlinearity, as in the case of initially monochromatic waves~\cite{Ricard2024}, although less efficiently. For higher amplitude solitons, this linear theory overestimates the observed localization length as nonlinear phenomena are not captured. 
Besides attenuating the leading soliton, we reported that a periodic lattice leads to the emergence of slower nondispersive pulses at each reflection of the lattice step whereas mostly dispersive waves are generated for the random lattice. Other lattice parameters could be varied in a longer canal such as the obstacle strength (bar height and width), or the randomness to investigate a continuous transition from the periodic ($\kappa=0$) to fully random ($\kappa=1$) cases. Finally, other phenomena could also be studied in such a setup, and analyzed using inverse scattering transform method~\cite{Gardner1967}, such as the possible existence of solitons within the band gap~\cite{Chen1987} or the existence of soliton gas in a periodic or random lattice~\cite{Suret2023}. The possible alteration of Anderson localization for other nonlinear systems, such as solitons governed by nonlinear finite-depth Schr{\"o}dinger equation, could also be investigated experimentally, and compared to theoretical predictions~\cite{MeiJFM2003}.



\begin{acknowledgments}
We thank Filip Novkoski for the scientific discussions. This work was supported by the Simons Foundation MPS No.~651463--Wave Turbulence (USA) and the French National Research Agency (ANR Sogood Project No.~ANR-21-CE30-0061-04 and ANR Lascaturb Project No.~ANR-23-CE30-0043-02).
\end{acknowledgments}

\end{document}